\documentclass[11pt]{article}
\usepackage[utf8]{inputenc}
\usepackage[left = 2.5cm, right = 2.5cm, top = 2.5cm, bottom= 2.5cm]{geometry}
 
\usepackage[colorlinks=true,urlcolor=blue,linkcolor=blue,citecolor=blue,breaklinks=true]{hyperref}
\usepackage{color}
\usepackage{float}
\usepackage{siunitx,booktabs,array}
\usepackage{fancyhdr}
\usepackage{ragged2e}
\usepackage{natbib}
\usepackage[none]{hyphenat}
\usepackage{graphicx}
\usepackage{wrapfig}
\usepackage{amsmath}
\usepackage{amsfonts}
\usepackage{csquotes}
\usepackage{xfrac}
\usepackage{arydshln}
\usepackage{nccmath}
\usepackage{multirow}
\usepackage{bm}
\usepackage{hhline}
\newcommand\mc{\multicolumn{1}{c}{}}
\newcommand\mcc{\multicolumn{1}{c|}{}}

\newcommand{\citationneeded}[1][]{{\color{blue} [citation needed]}}
\title{Unemployment Benefits and Job Quality: Unveiling the Complexities of Labour Market Dynamics}
\author{Jessica Reale, Frederik Banning, Michael Roos}

\begin{document}

\maketitle
\date{}

\begin{abstract}
\sloppy
    This study explores the impact of unemployment benefits on employment quality, job stability, and tenure within complex labour market dynamics. Given the macroeconomic consequences of changes in unemployment benefits, including their impact on employment rates and output growth, we develop a closed macroeconomic model that integrates heterogeneous households and adaptive firms and incorporates real-world entry-exit market mechanisms. The model considers personal values, social norms, and social network formation among workers as we examine the role of social contacts in mediating the effects of unemployment benefits on job-matching quality and labour market outcomes. We simulate the model across various scenarios where unemployment benefit schemes differ in level and/or duration.
    Our results suggest that extending the duration of unemployment benefits does not necessarily improve job-matching quality. Longer benefits may indeed reduce the effectiveness of social networks in job finding, indicating that social contacts play a key role in labour market dynamics.
\end{abstract}

\section{Introduction}
\sloppy
Economic downturns prompt governments to intervene with structural reforms and labour market policies aimed at increasing job security and unemployment insurance \citep{Galasso.2014}. As a result of job insecurity, businesses may experience lower profits and investments due to higher employee turnover, and households may reduce savings and consumption spending \citep{Cazes.2015}, prolonging recessionary episodes. The persistent levels of high unemployment characterising past (and present) economic crises sparked an extensive debate about unemployment insurance schemes (UI) and macroeconomic outcomes that revolve around two competing narratives: moral hazard vs. match quality.
Several empirical studies suggest that unemployment benefits reduce individuals' search efforts – \textit{moral-hazard} emphasis – and delay re-employment \citep[see][among others]{Meyer.1990, Katz.1990, Lalive.2006, Card.2007}. In this view, increasing benefits lead to  higher unemployment rates and slower recoveries from recessions \citep{OECD.2011}.\footnote{This interpretation embodies the trade-off between benefits and incentives such that individuals receiving economic support when unemployed might be more prone to prolong their time out of work, thereby deliberately delaying their exit from unemployment.} Conversely, other findings argue that increasing benefits levels and duration may enable job-seekers to look for better-matched employers  – \textit{match-quality} emphasis – through reduced liquidity pressures \citep{Belzil.2001, Tatsiramos.2009}. Longer unemployment spells may thus be beneficial since benefits ease financial difficulties, allow better job-matching outcomes, and promote employment stability and tenure \citep{Jackson.2013}. 
Recent research and policy discussions have increasingly recognised the importance of integrating job quality measures into evaluations of labour market policies \citep{Green.2020}. This recognition thus goes beyond monetary considerations as job quality is also determined by non-pecuniary factors that influence workers' well-being \citep{Cingano.2012}  and the role of unemployment benefits \citep{Howell.2011}.

Existing theoretical papers have made notable contributions to replicating labour market empirical regularities, mainly focusing on (i) workers' bargaining power \citep{Riccetti.2013}, (ii) wage determination mechanisms \citep{Dosi.2017}, (iii) skills deterioration \citep{Dosi.2018b}, and (iv) income inequalities \citep{Dosi.2018}. However, these analyses still tend to prioritise monetary aspects of labour market dynamics and overlook that the match quality between workers and firms significantly enhances employees' productivity within a specific organisational setting \citep{Jovanovic.1982, Failla.2017}, increasing job satisfaction and reducing labour turnover \citep{Farooq.2020}. 

Within the ongoing debate on unemployment insurance schemes and their influence on labour markets,  empirical evidence has yielded inconclusive results, largely attributed to the omission of certain critical features.
Indeed, current empirical analyses about the impact of UI on job quality report ambiguous evidence: either the effect is positive but weak, or it is negligible \citep{Tatsiramos.2014}. Moreover, job quality and stability exhibit time-dependent patterns: UI recipients who accept a job offer short before their benefits expire find themselves in lower-quality jobs \citep{Caliendo.2013}.
Liquidity constraints may thus affect individuals' behaviour in terms of job search and acceptance but not necessarily in terms of job quality. Existing studies link employment quality to objective labour market outcomes, i.e.,wages and hours worked \citep[e.g.,][]{Acemoglu.1999, Marimon.1999}. However, non-pecuniary factors are also determinant \citep{Bonoli.2104}.
The European Commission, for example, identified employees' skills, working conditions, job safety and satisfaction as additional determinants of job quality \citep{EC.2001}.
The rising awareness of the impact of job well-being on economic outcomes \citep[see][]{Muñoz.2011} thus led to the construction of job quality indexes that account for monetary, environmental and policy determinants simultaneously \citep{Cazes.2015}. The OECD job quality framework, indeed, features three key dimensions: (i) earnings quality\footnote{The quality of earnings is measured in terms of homogeneous preferences of income inequality.}, (ii)  labour market security, i.e.,unemployment benefits, and (iii) quality of the working environment. When conceptualising workplace quality, the OECD recognises the importance of employee relationships and the degree of autonomy workers can experience within a firm. However, the constructed index has serious flaws due to its reliance on homogeneous assessments of individual preferences. First, subjective attitudes and values are deemed as being "not easily amenable to policy" \citep[][page 14]{Cazes.2015}, despite labour market policies strongly depending on country-specific cultural traits \citep{Alesina.2015}. Second, workers' reactions to autonomy at the workplace cannot be considered from a \textit{one-size-fits-all} perspective since they strictly depend on agents' personal values \citep{Roos.2022, Banning.2023}. Third, how the circulation of job-related information impact job quality and matching is still ignored, leaving aside the "social returns to finding a job"  \citep[][page 292]{Cingano.2012} arising from heterogeneous workplace relationships.

Individuals who are unemployed and subsequently find new employment have the opportunity to establish new contacts and build relationships with colleagues, which can be beneficial for future job search endeavours. This social return of building a network of employed contacts may thus reduce unemployment spells \citep[see][among others]{Korpi.2001, Galeotti.2014} and be more beneficial than mere monetary outcomes. 
Existing literature has long confirmed that job-seekers find new employment opportunities through friends and former colleagues \citep[see][among others]{Holzer.1988, Granovetter.1995, Calvo.2004, Caliendo.2011} and that the lack of employed contacts might prolong unemployment \citep{Morris.1992}. Sociologists argue that the more similar individuals are, the more likely they are to form close relationships \citep[e.g.,][]{Kandel.1978}. As a result, friendships homophily might be determinant for better worker-firm matching outcomes in the labour market. Empirical findings suggest that the impact of social contacts on the quality of re-employment depends on the degree of friendship homophily \citep{Cappellari.2015}. When workers have heterogeneous personal values they may have divergent degrees of friendship homophily \citep{Lonnqvist.2016}. As such, the inquiry about unemployment benefits and their impact on job quality, stability and tenure should be re-oriented towards more intricate employee behaviours and social networks. 
Also, current job quality indexes – i.e., the OECD job quality framework \citep{Cazes.2015} – still assume workers' homogeneous preferences for firms' organisational settings and ignore – contrary to empirical findings - that the quality of re-employment may strongly depend on the channel the unemployed use to find a new job \citep{Montgomery.1991} and the degree of similarity of workers' contacts \citep{Cappellari.2015}. 
Therefore, it is still unclear how heterogeneous agents forming individual workplace relations with distinct degrees of cultural similarity influence the unemployed's job search behaviours and mediate the impact of UI on subjective employment quality and job stability, impacting macroeconomic outcomes such as (un)employment rates and output growth. 

This paper aims to address this theoretical gap by examining the impact of unemployment benefits on subjective employment quality, job stability, and tenure, with a particular emphasis on job matching quality.
This paper joins the debate about unemployment benefits and job-matching quality outcomes by modelling complex labour market dynamics where heterogeneous households and adaptive firms interact through typical entry-exit labour market mechanisms, i.e.,hiring, firing and quitting. We build a closed macroeconomic model – agent-based stock-flow consistent \citep{Caiani.2016, Dosi.2018b} – featuring employee behaviour as dependent on personal values and social norms \citep{Roos.2022}, forming a social network with other colleagues and reacting to endogenous organisational settings shaped by management strategies such as monitoring and incentives \citep{Banning.2023}. By doing so, workers with certain personal values are more or less satisfied according to their individual preferences for autonomy and group or individual rewards. The degree of job satisfaction impacts productivity and their propensity to stay at the firm or leave. Moreover, firms can monitor employee behaviour and fire non-performing workers. Since workers form a social network while performing their job tasks, they can be more or less embedded within the company. Underperforming workers who are more (less) socially embedded have a lower (higher) probability of being fired as we account for a turnover contagion effect \citep{Rubenstein.2019}. Fired and quitting employees then join the pool of unemployed individuals, receive unemployment benefits from the government and search for jobs through friends and former colleagues in their network. This paper thus evaluates a subjective measure of job quality as dependent on individual values and preferences for certain management strategies. In this setting, we account for several scenarios by altering the level and duration of unemployment benefits to mimic the implementation of security-based fiscal policies.
Our theoretical analysis thus contributes to the literature about unemployment benefits, job quality and stability by modelling complex labour market dynamics focused on the social components of agents' interactions \citep{Neugart.2018}, where job quality is subjective and depends on personal preferences for workplace autonomy and pay-for-performance schemes. Moreover, it integrates these factors into a broader macroeconomic context, moving past traditional monetary-based analyses of unemployment benefits and their macroeconomic implications.

The outline of this paper is as follows. Section \ref{sec:model} presents the main characteristics of the model and the entry-exit protocols implemented. Section \ref{sec: results} explains the simulations and the experiments conducted and describes the main results. The last section concludes.
    
\section{The Model}\label{sec:model}
\begin{table}[t]
    \centering
    \resizebox{\textwidth}{!}{\begin{tabular}{lllllll}
    \toprule
    & Firms & Households & Bank & Government & Central bank & $\Sigma$\\
    \toprule

        Inventories & $+INV$  & & & & & $+INV$\\
        Loans & $- L $ & & $+L$ & & & 0 \\ 
        Deposits & & $+D_{h}$ & $-D$ & & & 0 \\
        Bills & & & $+Bb$& $-B$ & $+Bcb$&  0\\
        High powered money & & &$+H$ & & $-H$ & 0 \\
        Advances & & &$-A$ & & $+A$ & 0 \\
      \textit{Balance} & $-NWf$& $-NWh$ & $-NWb$  &$+ GD$ & 0 & $-INV$ \\ 
      \bottomrule
      $\Sigma$ & 0 & 0 & 0  & 0 & 0 & 0 \\
      \bottomrule
    \end{tabular}}
    \caption{Balance sheet matrix.}
    \label{tab: bs}
\end{table}

We extend the VEPABM framework \citep{Roos.2022, Banning.2023} where (i) financial incentives, motivational factors, and endogenous social norms shape employee behaviour and company performance, (ii) employees' personal values influence working time allocation between personal tasks, cooperation, and shirking activities, (iii) workers build a social network that shapes corporate culture, and (iv) adaptive firms dynamically adjust their management strategies, namely financial incentives and monitoring. Employees have heterogeneous personal values \citep{Schwartz.2012b} that influence their working decisions within the firm. We assume that workers can desire a certain degree of autonomy at the workplace (openness-to-change vs. conservatism) and/or have diverse preferences for financial incentives (self-enhancement vs. self-transcendence). These motivational factors affect how they allocate their working time between personal, cooperative and shirking activities within certain managerial frameworks. Firms can indeed choose the degree of monitoring and the type of pay-for-performance (PFP) schemes to implement and adjust their strategy over time based on company outcomes. However, the firm might not be able to perfectly steer the organisation towards the desirable outcomes as employees' interactions lead to the emergence of social norms – i.e., corporate culture – which influence behavioural choices and in turn the formation of relational ties within the firm.  Each agent receives rewards based on individual performance and/or cooperative activities. Shirking does not contribute to individual performance and, as such, is not rewarded. Instead, employees who are caught shirking more than accepted receive written warnings ($ww_{i,t}$) and are perceived by the management as "non-performing". Employees receiving warnings become \textit{dissatisfied} within their working environment and decrease their productive efforts.\footnote{For the complete set of model equations of the VEPABM framework, please reference to \cite{Banning.2023}.}

\begin{figure}[t]
    \centering
    \includegraphics[scale = 0.9]{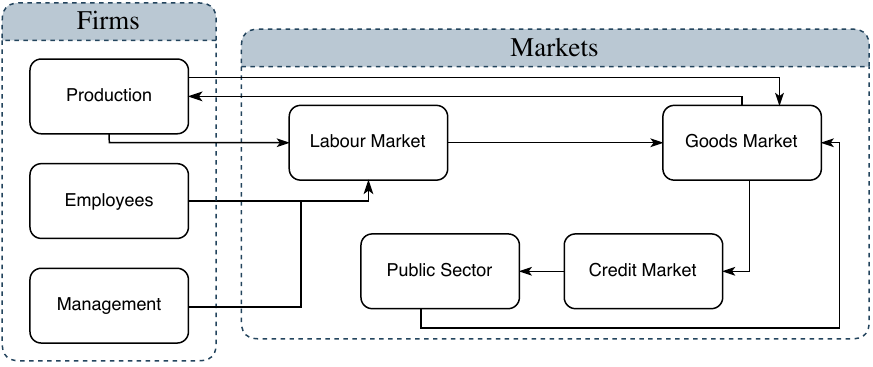}
    \caption{Model overview. Authors' own elaboration.}
    \label{fig:model-overview-main}
\end{figure}

These model characteristics are complemented by labour market processes like hiring, firing, and quitting within a broader macroeconomic context, as summarised in Figure \ref{fig:model-overview-main}.\footnote{A more complete overview of the model is devoted to Figure \ref{fig:model-overview} in the Appendix.} 
We thus build a hybrid agent-based stock-flow consistent model that features three aggregate sectors – the government, the central bank, and the banking sector – and heterogeneous firms and households. The balance sheet matrix at Table \ref{tab: bs} describes the stocks held  by  each  sector,  providing  a  snapshot  of  our aggregate framework. Firms $i \in N_{f}$ – where $N_{f} = [1, n_{f}]$ –  hold inventories $INV$, and ask for bank loans $L$. Households $i \in N_{hh}$ – with $N_{hh} = [1, n_{hh}]$ – 
hold deposits $Dh$, provided on demand by the banking sector. The bank holds high-powered money $H$, asks for advances ($A$) and buys a proportion ($Bb$) of government bills $B$ issued by the government. The central bank accommodates bank reserves and clears the bills market $Bcb$. We model a very simplified macroeconomic framework, in line with previous theoretical approaches \citep[e.g.,][]{Dosi.2018b}, to focus our attention on more realistic labour market practices and complex workers-employers dynamics. In this model, job (dis-)satisfaction impacts workers' retention probabilities, and firms have the ability to monitor and dismiss underperforming employees. The endogenous formation of social networks among colleagues affects workers' embeddedness within the company, influencing the likelihood of being fired and/or quitting the firm. When employees leave, they join the pool of unemployed individuals, receive government unemployment benefits, and utilise their network connections to search for new job opportunities. Firms rely on referrals in the hiring process by seeking recommendations from their most successful employees who have unemployed friends.
While household-workers are heterogeneous in their job-related behaviour, household-consumers are only indirectly affected by individual job performance, social norms, personal values and relational ties.
\subsection{Timeline}
In each simulation period, the following sequence of events takes place: \vspace{-0.7em}
\begin{enumerate}\setlength\itemsep{-0.1em}
    \item \textit{Price and output decisions}: firms set prices, make their production decisions by computing their output based on their expected sales  and inventories targets, and decide how much labour is needed for production;
    \item \textit{Hiring and firing}: firms can decide to fire non-performing workers and/or to hire new workers from the pool of unemployed agents. Moreover, when unemployed benefits expire, firms hire those agents who \textit{signal} themselves in the market; 
    \item\label{it: wage} \textit{Production of goods}: employees allocate their time between personal, cooperative and shirking activities based on personal values, social norms and their social network within the firm. Firms pay wages according to the PFP schemes implemented;
    \item \textit{Goods market}: households receive wages or unemployment benefits and pay taxes. The government and households make consumption decisions. If sales exceed firms' output and past inventories, we assume consumers are rationed homogeneously for the corresponding difference;

    \item \textit{Credit market}: the banking sector is assumed to be fully accommodative to firms' and households' demands. Firms ask for loans to finance changes in inventories. Banks compute their profits based on past period assets and liabilities. Households compute their disposable income based on wages, unemployment benefits, taxes and distributed firms' and banks' profits, and demand bank deposits;
    \item \textit{Bills market}: the central bank distributes profits to the government. The government collects households' taxes and issues bills, which are bought by the bank and, residually, by the central bank;
    \item \textit{Quitting}: less socially embedded employees can leave the firm after comparing wages and satisfaction levels with a  certain subset of peers.
\end{enumerate}

Since the aggregate behaviour of the system is quite standard in the SFC literature, we devote the full description of sectoral behavioural equations to Appendix \ref{SFC-eqs}. For the sake of the labour market protocol implemented in this model, only the aggregate productive sector's behaviour is covered in the following pages.
\subsection{Productive sector}
Firms set their prices (Eq. \ref{eq: prices}) following a mark-up rule on previous period unit costs, defined as the ratio of current period wage bill and output (Eq. \ref{eq: unit_costs}). 
Wages paid to employed workers depend on the rewards the management provides based on employees' performance and PFP schemes implemented (Eq. \ref{eq: wage}), following \cite{Banning.2023}.\footnote{Financial rewards are formalised as $R_{i,t} = \omega_{b} + \mu B_{i,t}$, where $B$ is the bonus  $B_{i,t} = (1-\lambda)O_{i,t} + \lambda (\frac{1}{n}) \sum_{j=1}^{n} O_{j,t}$ and depends on (i) incentive schemes, such that $\lambda$ equals to one (zero) identifies cooperative (individual) rewards, and (ii) workers' individual performance $O_{i,t} = \pi_{i,t} (p_{i,t}^{\phantom{i}(1-\kappa)}\cdot \bar{c}_{i,t}^{\phantom{i}\kappa})$, dependent on satisfaction-based productivity $\pi_{i,t}$, personal tasks $p_{i,t}$, average cooperation $\bar{c}_{i,t}$ and the exogenous degree of task interdependence $\kappa$.}
The wage bill (Eq. \ref{eq: wage_bill}) is the total rewards paid to employed workers $n_{i,t}$, determined by the labour-matching protocol explained in the following pages.
Every period, firms compute their expected sales (Eq. \ref{eq: sales-exp}) as a convex linear combination of their actual sales $s_{i}$ (Eq. \ref{eq: sales}) and past expectations $s^{exp}_{i}$.
Firms tend to achieve long-run and short-run inventory targets (Eq. \ref{eq: inv_target}, \ref{eq: inv_exp}).
Current output (Eq. \ref{eq: output}) is formalised as a Leontieff production function that takes the minimum between potential output (Eq. \ref{eq: potential_output}) and expected sales, the short-run inventory target and past period real inventories (Eq. \ref{eq: inv}). Potential output $y_{i,t}^{pot}$ depends on workers' time allocation as modelled in the VEPABM framework, that is $O_{i,t} = \pi_{i,t} (p_{i,t}^{\phantom{i}(1-\kappa)}\cdot \bar{c}_{i,t}^{\phantom{i}\kappa})$, with productivity $\pi_{i,t}$, individual activities $p_{i,t}$, average team cooperation $\bar{c}_{i,t}$ and an exogenous degree of task interdependence $\kappa$.\footnote{The degree of task interdependence captures the degree to which the performance of employee $i$ relies on the assistance of colleagues.} As such, production is demand-driven unless workers' individual output is not enough to cope with expected sales. When this happens, consumers are rationed homogeneously.
The flow of firms' inventories adjusts to the difference between output and actual sales, and nominal inventories are computed at current unit costs (Eq. \ref{eq: inventories}).
The amount of loans demanded  is defined as an accounting identity (Eq. \ref{eq: loans}), and firms' profits are distributed to households (Eq. \ref{eq: f_profits}).
\begin{align}
     & p_{i, t} = UC_{i,t-1}(1 + \bar{\rho}_{f}) \label{eq: prices}\\
     &UC_{i,t} = WB_{i,t}/y_{i,t}\label{eq: unit_costs}\\
     &  W_{i,t} =  \frac{\sum_{j \in n_{i, t}}R_{j,t}}{n_{i, t}}\label{eq: wage}\\
     &WB_{i, t} = W_{i, t}*n_{i,t}\label{eq: wage_bill}\\
     & s^{exp}_{i,t} = \bar{\beta}_{exp}* s_{i,t-1} + (1- \bar{\beta}_{exp})*(s_{i,t-1} - s^{exp}_{i,t-1})\label{eq: sales-exp}\\
     & s_{i,t} = \sum_{j \in N_{hh}}c_{j,t} + \bar{g}\label{eq: sales}\\
     &  inv^{T}_{i,t} = \sigma_{inv}*s^{exp}_{i,t}\label{eq: inv_target}\\
          &  inv^{exp}_{i,t} = inv_{i,t-1} + \bar{\lambda}_{exp}*(inv^{T}_{i,t} - inv_{i,t-1})\label{eq: inv_exp}\\
     & y_{i,t} = min(s^{exp}_{i,t} + inv^{exp}_{i,t} - inv_{i,t-1}, y_{i,t}^{pot})\label{eq: output}\\
     & y_{i,t}^{pot} = \sum_{j \in N_{hh}} O_{j,t}\label{eq: potential_output}\\
     &  inv_{i,t} =  inv_{i,t-1} + y_{i,t} - s_{i,t} \label{eq: inv}\\
     & INV_{i,t} = inv_{i,t} * UC_{i,t}\label{eq: inventories}\\
     & L_{i,t} = L_{i,t-1} + \Delta INV_{i,t}\label{eq: loans}\\
     &Pf_{i,t} = \sum_{j \in N_{hh}}C_{j,t} + G_{t} +  \Delta INV_{i,t} - WB_{i,t} - \bar{i}_{l}*L_{i,t-1}\label{eq: f_profits}
\end{align}
\subsection{Labour market}
\begin{figure}[t]
    \centering
    \includegraphics[scale = 0.63]{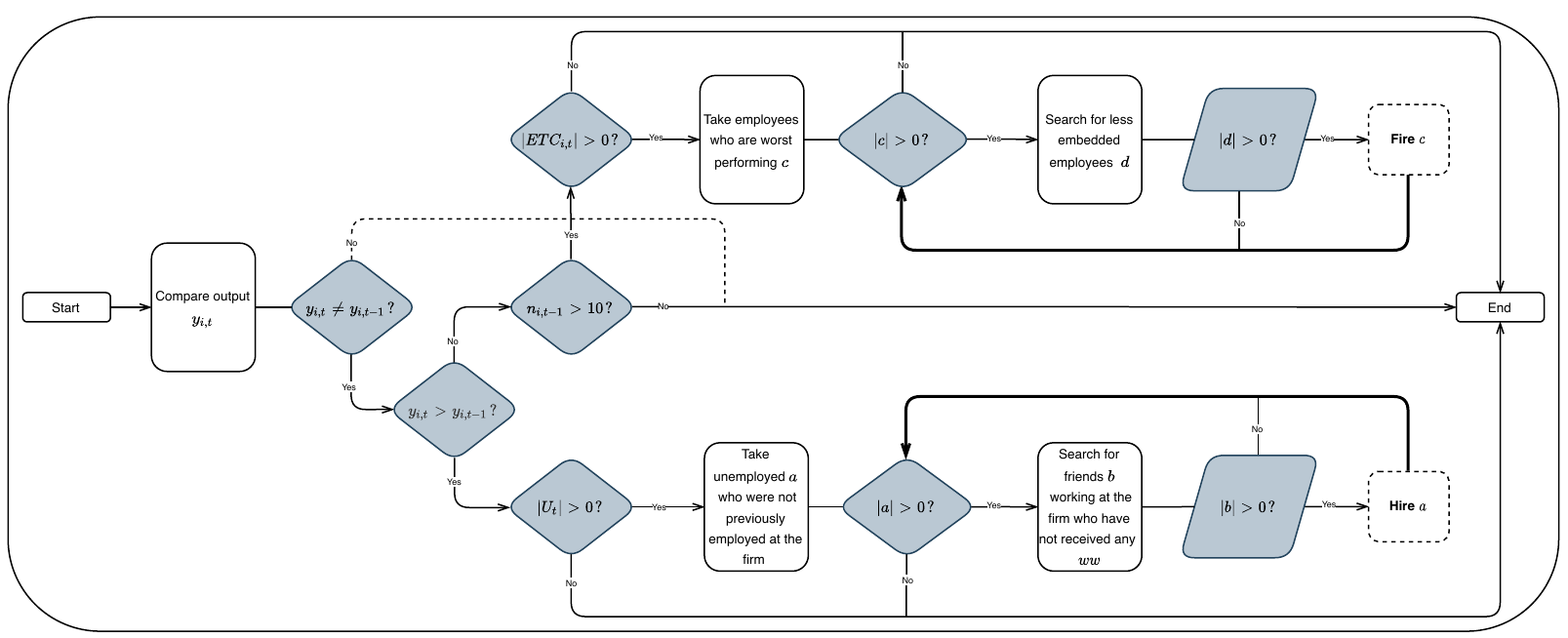}
    \caption{Hiring and firing protocol. Authors' elaboration.}
    \label{fig:lab_matching}
\end{figure}

The labour market protocol we implement is anchored to firms' change in output $\Delta y_{i,t}$ following \citep{Dosi.2017, Dosi.2018, Dosi.2018b}. The amount of workers to hire or fire depends on firms' demand for labour $n^{d}_{i}$ – see Eq. \ref{eq: lab_demand} – which is linked to current output $y_{i,t}$ and employees' average output $\bar{O}_{j,t}$, $\forall j \in n_{i, t-1}$. Therefore, any positive (negative) change in output leads firms to employ (deploy) 
$n$ unemployed households (workers), with $n = |n_{i,t}^{d} - n_{i,t-1}|$.
The labour matching protocol implemented is depicted in Figure \ref{fig:lab_matching}.

\begin{equation}\label{eq: lab_demand}
    n^{d}_{i,t} =  \frac{y_{i,t}}{\bar{O}_{j,t}}
\end{equation} 

When faced with an increased change of output $\Delta y_{i,t} > 0$, firm $j$ initiates the process of searching for suitable candidates within the pool of unemployed individuals $U_{t}$ it has not previously employed, see Eq. \ref{eq: unemployed-pool}
\begin{equation} \label{eq: unemployed-pool}
U_{t} = \{i : i \in N_{hh}  \;\land\;  i\notin n_{j,t - 1}\}
\end{equation}

\noindent Embeddedness research suggests that firms tend to employ referrals from current employees as an effective recruitment strategy \citep{Rubenstein.2019}. Thus, in our model, we assign a certain number of friends to each agent and we assume that friendships form a \textit{stable} personal network which does not change over the simulation periods. While friendships might be "flexible" due to geographical distance and/or personal development \citep{Becker.2009}, we support the notion that as individuals enter the age range suitable for employment, there is a tendency for their friendships to become more stable over time \citep{Poulin.2010, Granic.2015}. 
These friends serve as potential referrers during the hiring and firing process, reflecting managers' attitude to consider recommendations from current employees \citep{Fernandez.1997, Castilla.2005, Breaugh.2013}.
The selection of friends in our model is influenced by agents' individual value types, which shape their social interactions \citep{Lonnqvist.2016}. 
For instance, individuals with self-enhanced value orientations are less likely to form friendships with others who share similar values due to the potential for rivalry. On the other hand, conservative households with deeply ingrained conservative traditions may exhibit a low probability of forming friendships with individuals from the same value group, given substantial variations in conservative beliefs among different individuals. By incorporating these dynamics, our model captures the complexities of social interactions and the influence of personal values on friendship formation.
Friends thus function as referrers for unemployed households since managers would benefit from hiring job candidates with a higher propensity to be embedded in the working context \citep{Allen.2013}. Therefore, firm $j$ will hire workers $a$ – see Eq. \ref{eq: a} – having at least one friend $b$ working at the firm who has not received any written warning during the previous period or who has not received more than three warnings, Eq. \ref{eq: b}. 
The motivation behind this restriction lies on the so-called "referrer's performance effect" \citep{Yakubovich.2006}, which is based on empirical evidence showing that the advantage of a referral increases alongside the internal referrer's performance.
\begin{align}
     &a=\{i:i \in U\;  \land\;  i \notin n_{j,t-1}\} \label{eq: a}\\
     &b = \{j: j \in F_{i}\;  \land\;  (|ww_{j, t-1}| = 0\; \lor\; |ww_{j,t}| < 3), \forall i,j \in N_{hh} \; \land \; j \ne i \}\label{eq: b}
\end{align}

When firms experience decreasing output (i.e. $\Delta y_{t}< 0$), managers start firing non-performing workers ($i \in ETC_{t}$), if any. Firm $j$ thus looks at those employees who received at least three written warnings throughout their employment contract or whose last warning was received during the previous period, see Eq. \ref{eq: c}. However, the decision to fire such workers is influenced by the level of social embedding within the firm.

\begin{figure}[t]
    \centering
    \includegraphics[scale = 0.6]{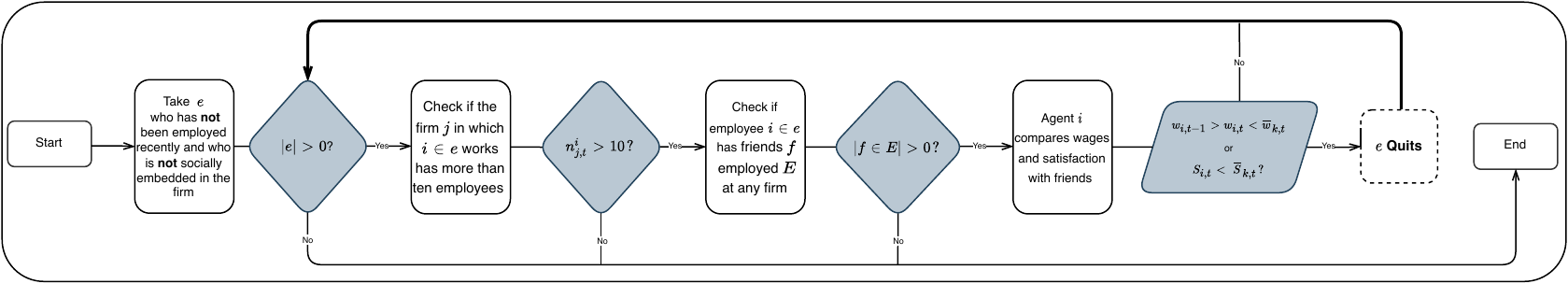}
    \caption{Quitting protocol. Authors' elaboration.}
    \label{fig:quit}
\end{figure}
\begin{equation}\label{eq: c}
     c=\{i:i \in ETC_{t}\;  \land \; |ww_{i,t}| \ge 3  \;\land\;  ww_{i,t}  = t-1\} 
\end{equation}
According to \cite{Dalton.1977}, firms find it challenging and unproductive to fire those employees who are well-socially embedded within the firm. The negative consequences of firing such employees may extend beyond their individual well-being and impact the remaining workers through a "turnover contagion effect" \citep{Rubenstein.2019}. As such, employers have a certain degree of embedding pressures to retain certain employees. We thus allow the management to search for those non-performing workers who are also less socially involved within the company, i.e., those having a lower interaction intensity than the mean of the other observed workers, Eq \ref{eq: d}. Therefore, firms must be able to strike a balance between performance needs and the preservation of social cohesion.

\begin{equation}\label{eq: d}
    d = \{i:i \in ETC_{t}\; \land\; \bar{e}_{i,j,t} < \sum_{i\in N_{t-1}}  \bar{e}_{i,t}, i \neq j\}
\end{equation}

Our model also incorporates voluntary turnover, allowing employees to decide whether to continue working for the firm or seek alternative employment opportunities.
Every period, workers who have been employed at firm $j$ in the previous two steps can decide to leave the company (Eq. \ref{eq: e}), see Figure \ref{fig:quit}. The 2-step time frame mimics the behaviour of "urgent leavers", for whom low job satisfaction and low commitment induce high pressures for turnover \citep{Van.2008}.
\begin{equation}\label{eq: e}
    e = \{ i : i \in n_{j,t} \;\land\; i \in n_{j,t-1}, \forall j \in N_{f} \}
\end{equation}
Employees compare their wage and satisfaction levels with friends who are employed, whatever the company, acting as endogenous reference groups which differ among individuals according to their value-based personal network formation \citep{Hauret.2019}, Eq. \ref{eq: f}.
\begin{equation}\label{eq: f}
    f = \{k : k \in F{i} \; \land\; k \in n^{i}_{j,t}, \forall j \in N_{f}\; \land\; \forall i \in e  \}
\end{equation}
They quit the firm if (i) their wage bill has decreased – $w_{i,t} < w_{i,t-1}$ – and (iia) is lower than the one earned by friends $\bar{w}_{k,t}$, or (iib) they are less satisfied than their friends, i.e.,$S_{i,t} < \bar{S}_{k,t}$.\footnote{Please note that $\bar{w}_{k,t} = \frac{\sum_{k\in f} w_{k,t}}{|f|}$ and $\bar{S}_{k,t} = \frac{\sum_{k \in f}S_{k,t}}{|f|}$.} By doing so, we capture the influence of social comparisons and the importance of social connections at the workplace \citep{Fischer.2009}. Workers who are more socially connected and embedded within the firm tend to be more reluctant to quit their jobs, as embeddedness theory suggests. Indeed, individuals with stronger social ties within an organisation are more likely to develop a sense of commitment and attachment, making them less likely to leave voluntarily. In this model, social cohesion at the workplace thus acts as an additional factor influencing employees' decisions regarding job retention.
\subsection{Unemployment benefits}
Unemployed households receive fiscal financial support. 
The government provides a dole as a proportion of the past individual wage received when last employed at time $x$ (Eq. \ref{eq: un_benefits1}). We also assume the existence of a standard $60\%$ poverty line \citep{Almeida.2022}, computed over the median of current wages paid to employed households $\widetilde{W}_{t}$, such to guarantee a minimum consumption expenditure. Unemployment benefits are assumed to expire after a certain period of time $\epsilon$ depending on the scenario implemented (see second row of Table \ref{tab:scenarios}). Upon expiration of benefits, unemployed individuals \textit{signal} on the labour market that they are making a concerted effort to find employment, and if unsuccessful, receive the minimum dole set by poverty levels ($\delta^{p}$). 
\begin{align}
     &UB_{i,t} = 
      \left \{ \begin{array}{ll}
      \text{max}(\bar{\delta}_{g}\cdot W_{i, t-x}, \bar{\delta}^{p}\cdot  \widetilde{W}_{t}) & \text{if}\;\; t < \epsilon \\
       \bar{\delta}^{p}\cdot  \widetilde{W}_{t}   &\text{otherwise} 
     \end{array} \right.  \label{eq: un_benefits1}
 \end{align}
The parameter $\bar{\delta}_{g}$ is the so-called "replacement rate", which conveys the government's attitude towards stability-based fiscal policies. We implement nine scenarios by altering the \textit{level} of the replacement rate and the \textit{maximum duration} ($\epsilon$) of benefits. As such, we only focus on the generosity of fiscal benefits, leaving aside the other two constituting elements of UI, namely entitlement conditions and the strictness of job-search requirements. We account for the 2019 median values of two advanced economies, Belgium and the United Kingdom, reporting a 18-months vs. a 6-months maximum duration \citep{Asenjo.2019}. Along the same lines, we assign the high and low values of $\bar{\delta}_{g}$ from the "net replacement rate in unemployment" measure obtained from the 2023 OECD database for the same two countries \citep{OECD.2023}. 
These factors yields four distinct scenarios based on varying levels and maximum durations: High vs. Low, characterised by a $69\%$ and $35\%$ replacement rate, and Long vs. Short, featuring benefits ending respectively after $540$ and $180$ periods.\footnote{Please note that one period corresponds to one working day.} The combination of $\delta_{g}$ and $\epsilon$ results in four additional mixed scenarios that allow us to assess how the postulated economy performs when the amount and the duration of benefits interact.
For the sake of comparability, we also set a theoretical baseline scenario where $\delta_{g} = 0.52$, i.e., the average between the two, and a maximum duration of one year ($360$ periods).
\begin{table}[t]
    \centering
    \begin{tabular}{|c|c||cc||cc||cccc|}
    \hhline{~~|--------}
    \mc& \mcc & \multicolumn{2}{c||}{Levels $\delta_{g}$} & \multicolumn{2}{c||}{Duration $\epsilon$} & \multicolumn{4}{c|}{Levels $\delta_{g}$ and Duration $\epsilon$}\\
    \hhline{~|-|--||--||----}
\mcc & \textbf{Baseline}  & High & Low  & Long & Short & High-Long & High-Short & Low-Long & Low-Short\\
\hline \hline
    $\delta_{g}$ &$\bm{0.52}$ & $0.69$&$0.35$ &$\bm{0.52}$ &$\bm{0.52}$ & $0.69$&$0.69$ &$0.35$ &$0.35$\\ \hline
    $\epsilon$  &$\bm{360}$ &$\bm{360}$ & $\bm{360}$& $540$&$180$ &$540$ &$180$ &$540$ &$180$\\
    \hline
    \end{tabular}
    \caption{Scenarios implemented.}
    \label{tab:scenarios}
\end{table}

\section{Simulations and results}\label{sec: results}
We simulate the model over $1080$ periods (i.e., twice the maximum $\epsilon$ considered), and we perform $100$ parallel replicates for each of the nine scenarios at table \ref{tab:scenarios}. The computation of the initial values of the model can be found in the Appendix. We report the results after having excluded the first $50$ steps (burn-in period) and de-trended the time series using the Hodrick-Prescott filter for quarterly data to have a clearer understanding of the dynamics of the model. 
\begin{figure}[t]
    \centering
    \includegraphics[scale = 0.5]{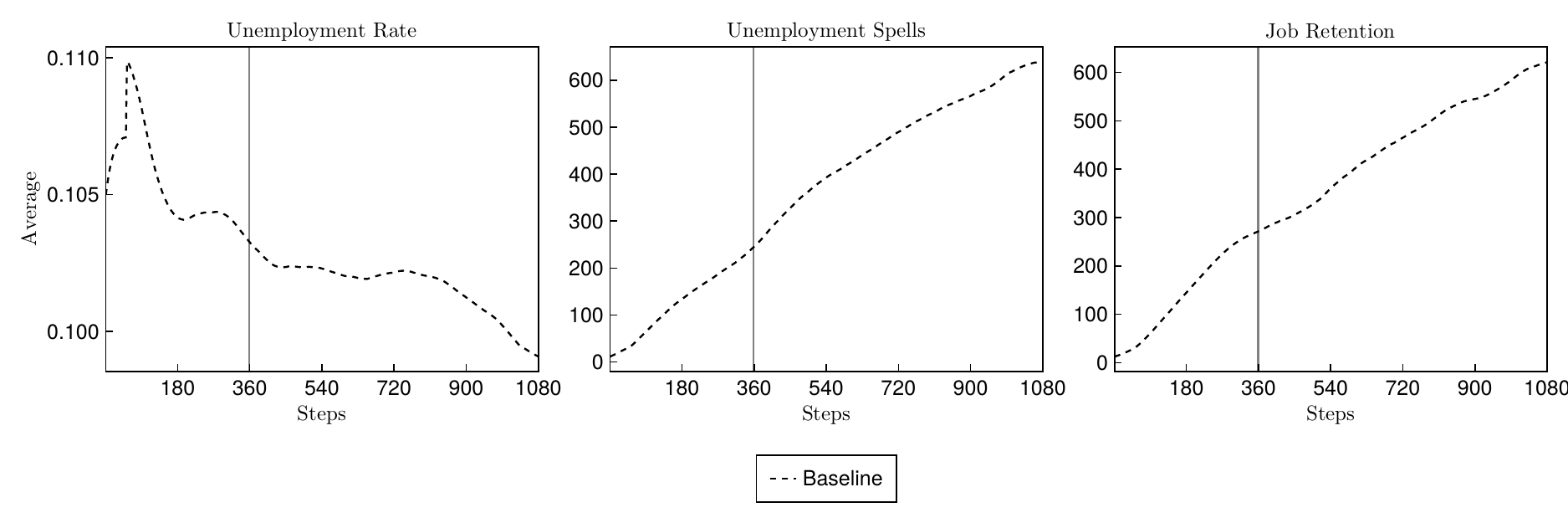}
    \caption{Labour market dynamics – Baseline scenario. The vertical line refers to the expiration date of benefits $\epsilon$.}
    \label{fig:lab_market-base}
    \includegraphics[scale = 0.5]{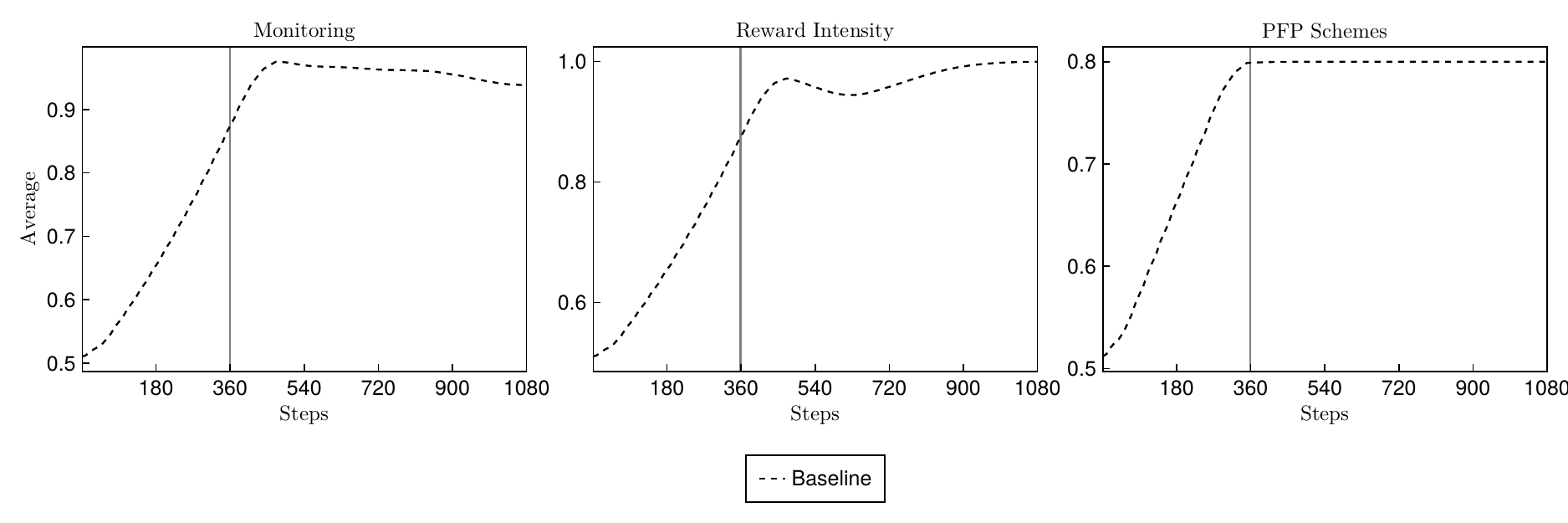}
    \caption{Management orientation – Baseline scenario. The vertical line refers to the expiration date of benefits $\epsilon$.}
    \label{fig:management-base}
\end{figure}

\begin{figure}[t]
    \centering
    \includegraphics[scale = 0.5]{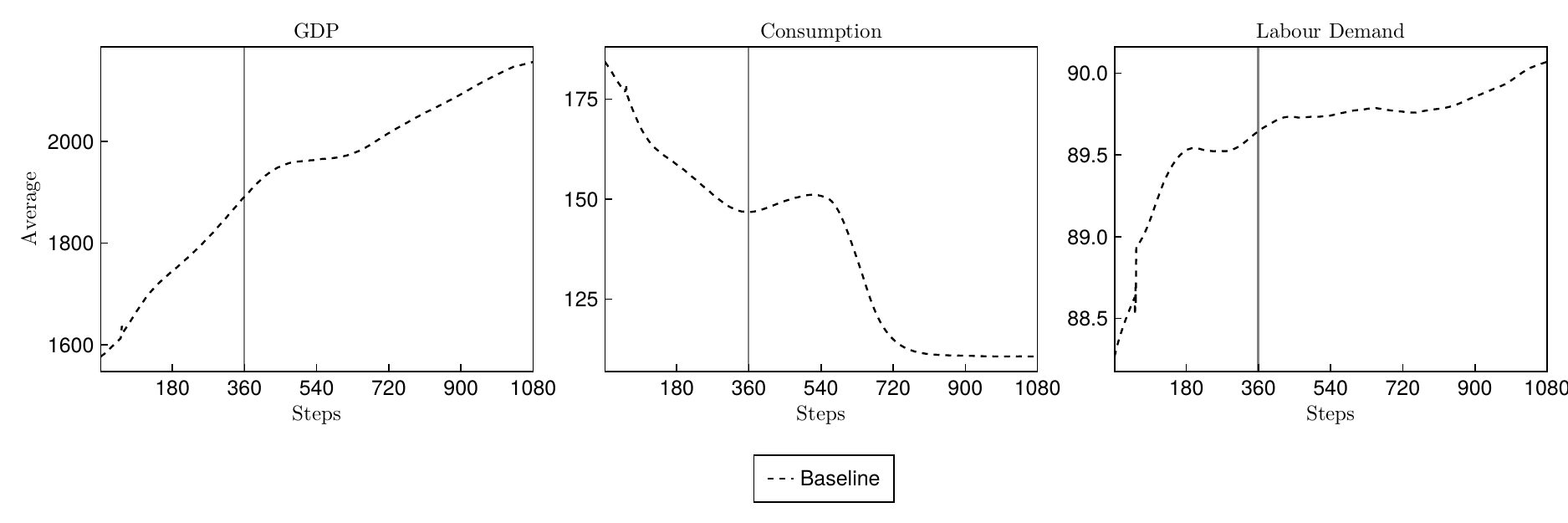}
    \caption{Real economy. Vertical lines refer to the expiration date of benefits $\epsilon$ (baseline scenario in bold).}
    \label{fig:real_economy-base}
\end{figure}

Figure \ref{fig:lab_market-base} reports the dynamics of the labour market for the baseline setting only. The unemployment rate (left plot) is generally low despite the restrictions imposed in the hiring process and tends to zero towards the end of the simulation period. This dynamic is triggered by increasing demand for labour (Figure \ref{fig:real_economy}) and the signalling mechanism of unemployed whose benefits expire long before the end of the simulation period. Indeed, when benefits end at period $360$ (bold grey line), the proportion of people who stay unemployed quickly decreases.
The duration of unemployment (employment) measure gets incremented by $1$ for each step in which households stay in unemployment (employment) and gets reset to $0$ when they find a suitable job (when they are fired or quit the company).
The central (right) plot of Figure \ref{fig:lab_market-base} suggests that the normalised mean duration of unemployment gradually increases and reaches its peak at the end of the simulation period at around $0.5$, i.e., in the end, agents spend $50\%$ of the total time available in unemployment.
The decreasing unemployment rate mixed with increasing labour demand (Figure \ref{fig:real_economy-base}) suggests an overall high level of job retention, which is confirmed by the downward trend of labour turnover (left plot of Figure \ref{fig:lab_market}).
This is typically viewed as a positive sign of employee satisfaction, job security, or alignment with company goals.

\begin{figure}[t]
    \centering
    \includegraphics[scale = 0.5]{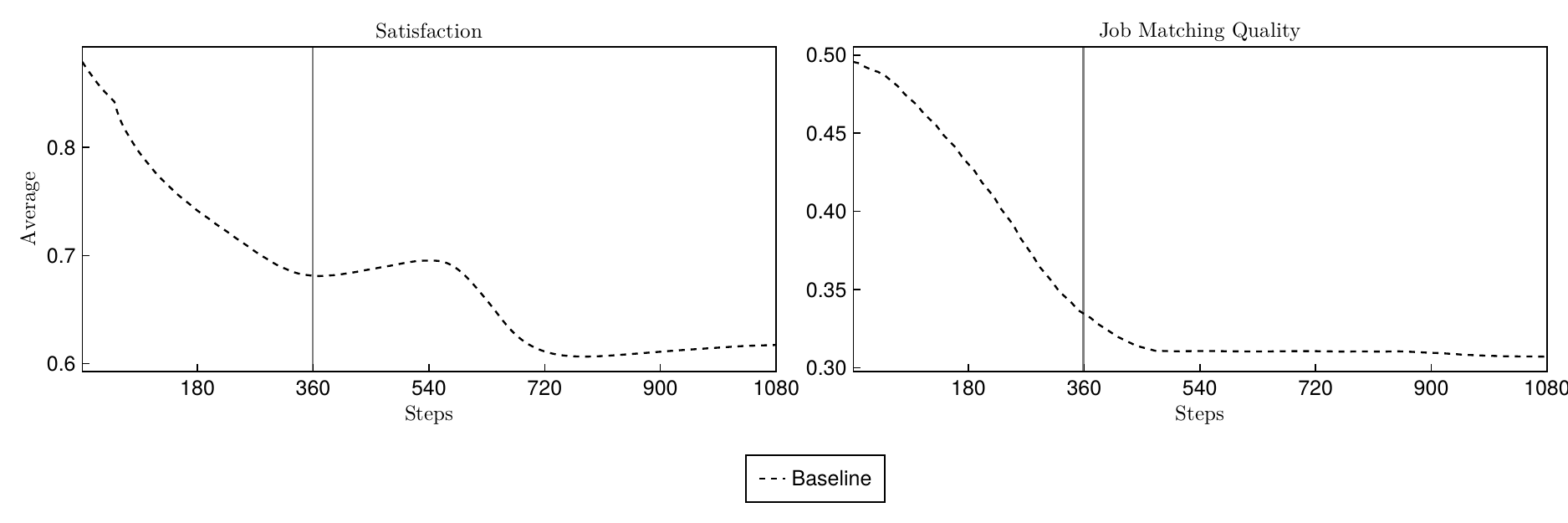}
    \includegraphics[scale = 0.5]{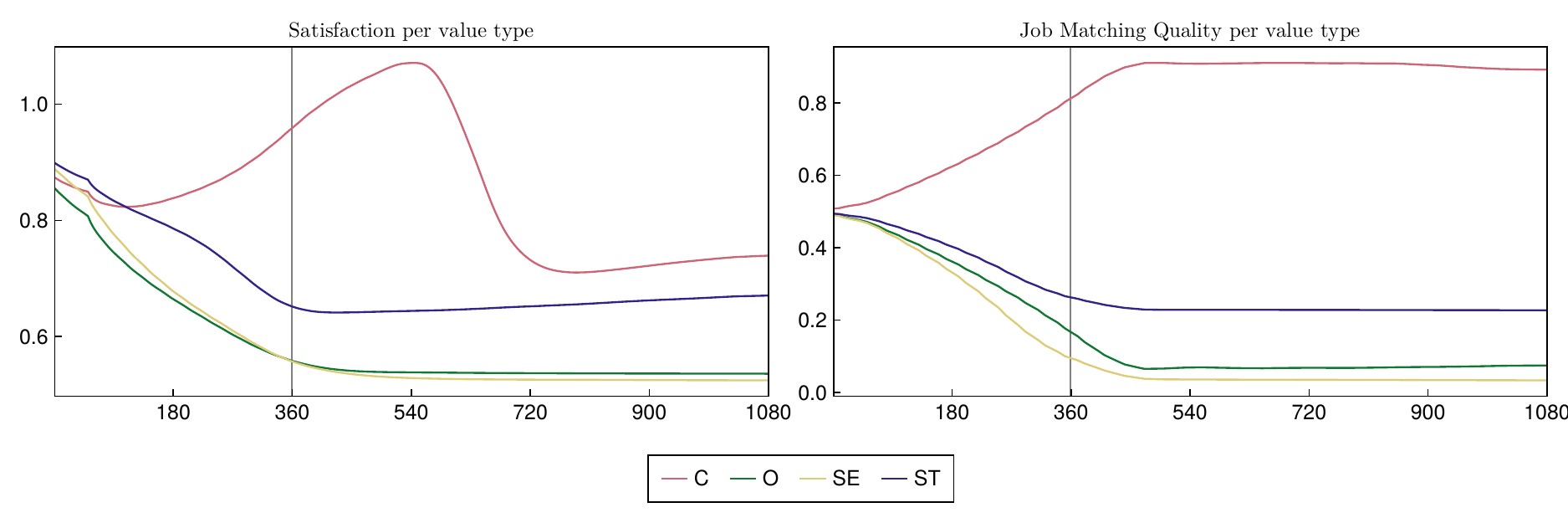}
    \caption{Workers' productivity and job quality – Baseline scenario. The vertical line refers to the expiration date of benefits $\epsilon$.}    \label{fig:workers-base}

\end{figure}

However, employees' satisfaction, as reflected in their productivity (see Figure \ref{fig:workers-base}), consistently decreases, potentially raising contradictions in the analysis.
To better understand the link between labour turnover and agents' satisfaction, we must look at how each agent value type reacts to companies' very high monitoring efforts (Figure \ref{fig:management-base}).
The bottom panel of Figure \ref{fig:workers-base} depicts productivity and job matching quality per value type.
Job matching quality is defined as the mean base satisfaction $S^0$ for each agent $i \in N_{hh}$.
It is assumed that this variable can be used as a proxy because it is updated every time an agent gets hired by another firm as well as when the management strategy changes.
As such, it can be seen as a momentary snapshot of agents' person-organisation fit, i.e. the long-run quality of their current job match without the temporary influences from recent experiences (like receiving a written warning).
As expected, the missing inverse link between job retention and satisfaction is driven by the downward trend of the majority of workers of $O$, $SE$ and $ST$ values.
Agents $C$ satisfaction exhibits an inverse U-shaped curve stabilising at around $0.75$, above the other agent types.
Given the high level of overall monitoring, newly employed $C$ agents also experience an increasing trend of job quality, which then converges to almost the maximum level, opposite to the other value types.
Furthermore, agents seem to be more concerned with monitoring efforts than the type of PFP schemes implemented.
Companies' cooperative orientation – reflected by PFP stabilising at $0.8$ – favours ST-type employees (blue line), but not enough to offset the productivity and quality effects of low freedom within a company. 

Overall, the evolution of the management strategy mimics the findings in \cite{Banning.2023} that the management’s behaviour is only imperfectly aligned with the social dynamics within the firm. As a result, only one of the four value-types (C-agents) reaches a high satisfaction level. Especially O-type agents and SE-type agents are rather dissatisfied impairing their performance and hence output.

\subsection{Scenarios comparison}

\begin{figure}[t]
    \centering
    \includegraphics[scale = 0.5]{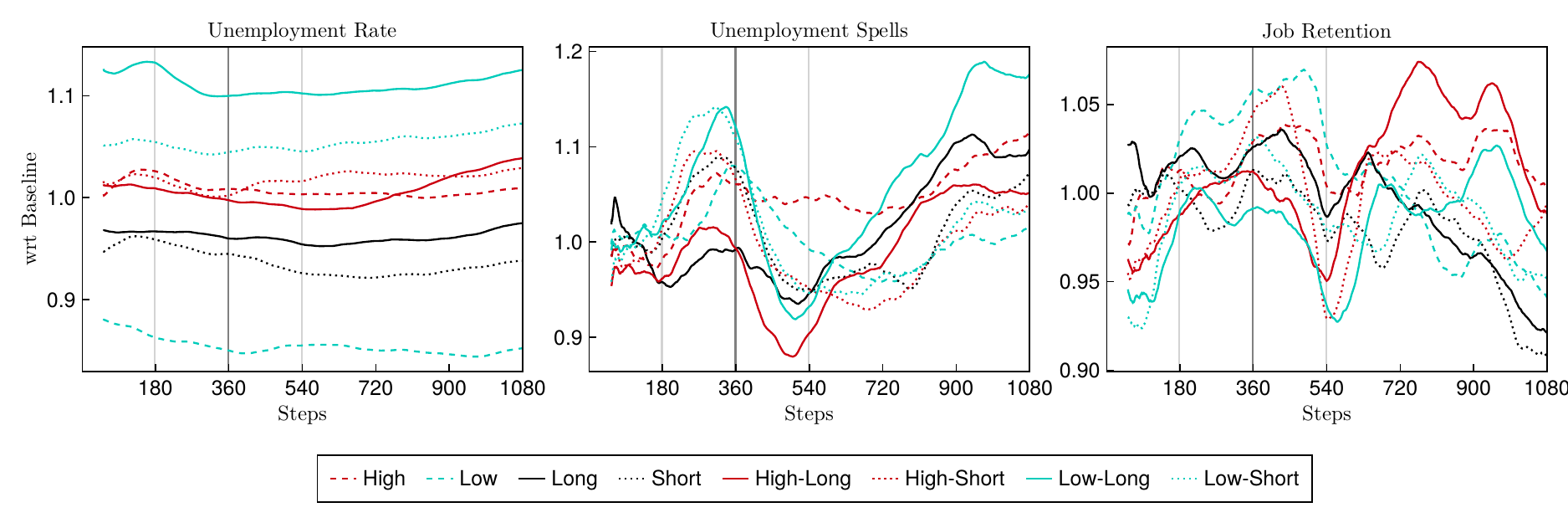}
    \caption{Labour market dynamics. Vertical lines refer to the duration of benefits $\epsilon$ (baseline scenario in bold).}
    \label{fig:lab_market}
\end{figure}

Figure \ref{fig:lab_market} shows the labour market dynamics in comparison to the baseline case for the eight scenarios.
In general, the effects of the different unemployment benefit schemes on the unemployment rate (left panel) are quite intuitive.
High benefits lead to higher unemployment than low benefits.
Similarly, unemployment is high, when the duration of the benefits is long, and low when the duration is short.
However, combining the two dimensions does not lead to the results one might expect: the unemployment rate is higher for low-short than for high-long.
From the perspective of individuals' incentives to accept new jobs, we might expect that the reverse is true.
The highest unemployment rate is obtained in the low-long scenario, while the lowest unemployment rate results in the low scenario.

While the unemployment rate mainly differs with regard to the level in the different scenarios, the unemployment spells exhibit interesting dynamics.
There are roughly three different patterns in the unemployment spells.
The most obvious pattern is that the unemployment spells in the scenarios long, high-long and low-long are correlated.
They show a peak before step 360, followed by a rather strong decline before step 540 and then a steady increase reaching a relatively constant level after step 900.
The second pattern is the correlation of the scenarios short, high-short and low-short. 
While the unemployment spells in those scenarios also peak around step 360, both the following decline and the increase are less pronounced than in the first group. 
Finally, the unemployment spells in scenario high increase from the beginning, but remain relatively high after step 360 with no strong decline. 
In scenario low, there is a decline, but no strong increase afterwards. 
This suggests that modifying the duration of the unemployment benefits adds volatility to the unemployment spells. 
The high unemployment rate in scenario low-long corresponds to the high numbers of unemployment spells. 
The dynamics of job retention are qualitatively quite similar in all scenarios.

\begin{figure}[t]
    \centering
    \includegraphics[scale = 0.5]{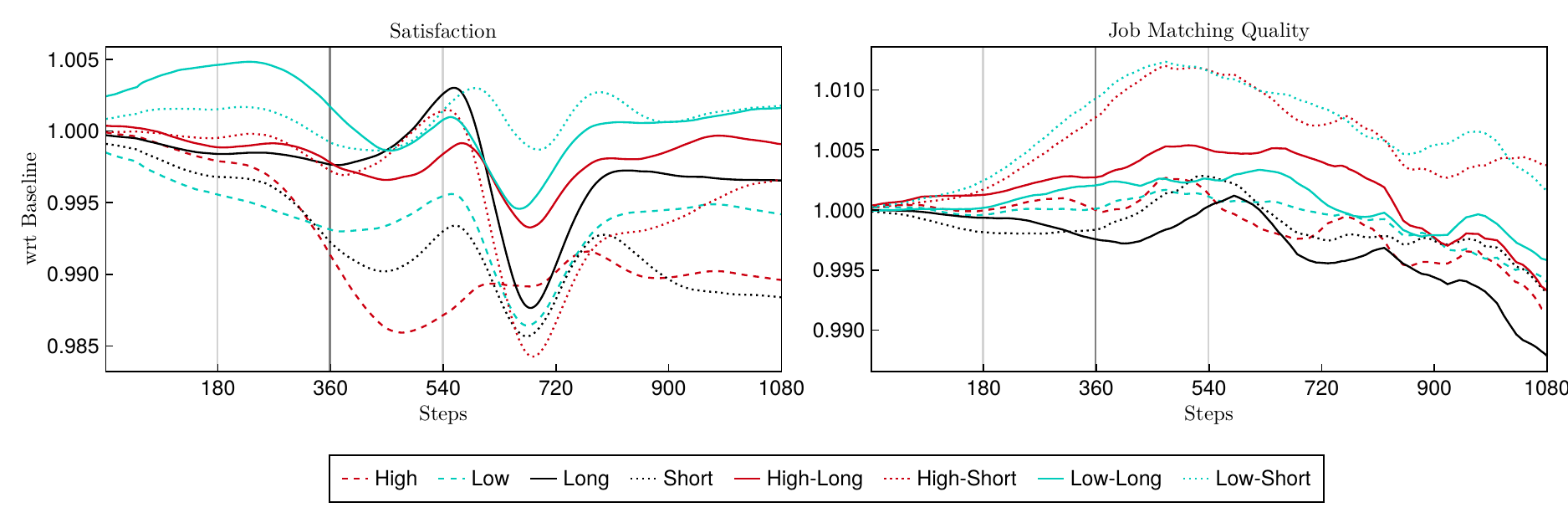}
    \caption{Workers' satisfaction and job quality. Vertical lines refer to the duration of benefits $\epsilon$ (baseline scenario in bold).}
    \label{fig:workers}
\end{figure}

Figure \ref{fig:workers} shows two noteworthy findings.
First, in all scenarios the drop in job satisfaction around step 700 already observed in Figure \ref{fig:workers-base} is present.
This means that the unemployment benefit scheme does not alter the internal dynamics of the firms, which are driven by the management’s strategy adjustments.
In terms of satisfaction levels, the differences between the scenarios and the baseline are small.
The second remarkable finding is that job matching quality is highest in the scenarios high-short and low-short, i.e. it seems to be driven by short durations of unemployment benefit payments.
This is confirmed by the low level of job matching quality in scenario long, especially in the second half of the simulation.

\begin{figure}[t]
    \centering
    \includegraphics[scale = 0.5]{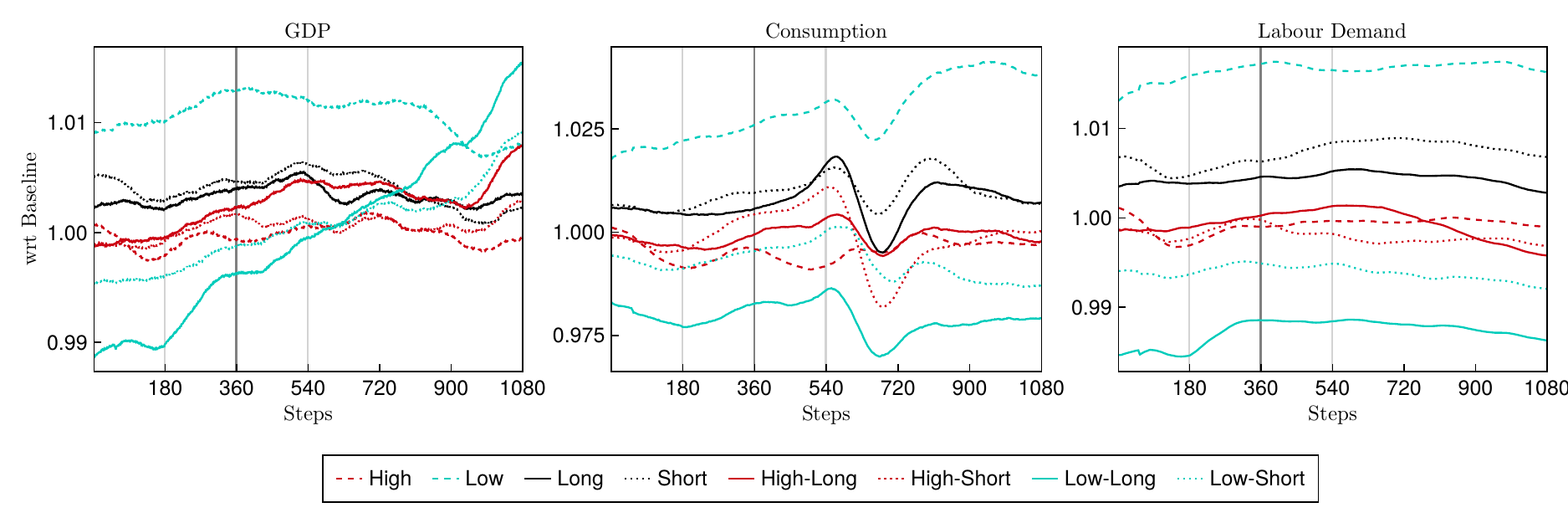}
    \caption{Dynamics of the real economy. Vertical lines refer to the duration of benefits $\epsilon$ (baseline scenario in bold).}
    \label{fig:real_economy}
\end{figure}

The macroeconomic effects of the different unemployment schemes are shown in Figure \ref{fig:real_economy}.
In most scenarios, GDP is quite similar (see left panel), and only two scenarios differ.
In scenario low-long, GDP initially is lowest, but then increases steadily until it is highest at the end of the simulation period.
In scenario low, GDP is higher than in the other scenarios throughout most of the simulation period and only in the end GDP declines slightly.
Note that these two scenarios are also the scenarios with extreme unemployment rates (lowest unemployment rate in low and highest in low-long).
Scenario low-short shows a similar GDP pattern to low-long but at a higher level. Consistently, this scenario exhibits the second-highest unemployment rate. 

Aggregate consumption in the middle panel of Figure \ref{fig:real_economy} behaves in a way that fits both the unemployment rate and GDP.
Scenario low-long, which has the highest unemployment rate shows the lowest level of consumption, which is plausible.
Analogously, consumption is highest in scenario low, which has the highest GDP and the lowest unemployment rate.
Again, the other scenarios are quite similar and do not differ much from the baseline scenario.
An interesting occurrence is the dip in the consumption time series shortly before step 720, which is present in all scenarios.
This fits the drop in the satisfaction level of C-type agents (see Figure \ref{fig:workers-base} and Figure \ref{fig:workers}) and hence some quits of these agents. 
The graphs of labour demand (right panel of Figure \ref{fig:real_economy}) look very similar to the consumption graphs except for the missing decline close to step 720. 

\section{Conclusions}

This paper presents a stock-flow-consistent agent-based macroeconomic model, similar to other models in this literature.
The novelty of this model is its realistic treatment of the labour market and agents’ behaviour on the job, which is driven by their values.
In addition, firms adjust their management strategies by changing the level of monitoring workers, performance premia and the degree to which group performance is rewarded.
A particularly realistic feature of the model, supported by a host of empirical evidence, is the informal hiring mechanism that relies on the recommendation of potential new workers from the employees’ friendship network. 

In this paper, we focus on the analysis of different unemployment benefit schemes on macroeconomic variables which is a topic that has been studied before using other models.
Our analysis confirms previous findings that low unemployment benefits reduce the unemployment rate, while higher benefits lead to higher unemployment given that the duration of the benefit payment in both cases is intermediate.
However, we show that the interaction of the level of the benefits with the duration of the payment is important in our model.
Interestingly, both for shorter and longer durations than in our baseline, the unemployment rate increases relative to the baseline if they are combined with a low level of benefits.
In general, longer durations seem to increase the unemployment rate and lead to strongly increasing unemployment spells in the second half of the simulation period.
In contrast to what is discussed in the literature, longer payment of unemployment benefits does not increase the job-matching quality in our model.
On the contrary, short durations either combined with high benefits or low benefits increase the job matching quality.

One reason for this finding might be that the recommendation mechanism might work less well if unemployment benefits are paid for a longer time.
As we have seen, the unemployment rate is high in scenario low-long and labour demand is low. 
The unemployment rate is also high in scenarios high-short and low-short, but in these scenarios, the unemployment spells are relatively short. 
Long unemployment spells make it more likely that friends have lost or changed their jobs, making it more difficult for an unemployed person to find a well-matching job via their friendship networks. 
Further analyses have to show whether this hypothesis is correct. 
If it turns out to be correct, our paper makes a strong case for the importance of realistic modelling of the labour market.

\bibliographystyle{apalike} 
\bibliography{ms}

\newpage
\appendix
\section{Sectoral behavioural equations}\label{SFC-eqs}
\setcounter{equation}{0}
The government buys a fixed proportion $\bar{g}$ of firms' output at the current price level (Eq. \ref{eq: gov_spending}) and collects taxes on households' previous wage bills (Eq. \ref{eq: gov_taxes}). New bills are issued as residuals through the government's accounting identity. 
The bank provides deposits to households on demand (Eq. \ref{eq: bank_deposits}), asks for the central bank's reserves as a fixed proportion of deposits (Eq. \ref{eq: hpm}), and buys government bills residually (Eq. \ref{eq: bank_bills}). If bills are null, banks' advances act as a residual variable (Eq. \ref{eq: bank_advances}).
Banks' profits are defined through an accounting identity and are assumed to be distributed to households.
The central bank clears the bills market (Eq. \ref{eq: cb_bills}), and distributes profits to the government (Eq. \ref{eq: cb_profits}), and provides advances to banks on demand.
Households consume a proportion of past disposable income $Yd_{i,t}$ and past deposits $Dh_{i,t}$ (Eq. \ref{eq: cons}) (standard in PK literature), and pay taxes on previous wages, see Eq \ref{eq: hh_taxes}. The nominal value of consumption is expressed at the current price level (Eq. \ref{eq: real_cons}).\footnote{We denote with $p^{*}_{t}$ the price set by the matched consumption firm.}
The amount of deposits held is defined through accounting identity. 
The redundant equation, i.e.,the accounting identity left out from the model equations, is the one related to the central bank's capital account $H_{t} = Bcb_{t} + A_{t}$.
\begin{align}
     & G_{t} = \bar{g}*\sum_{i \in N_{f}}p_{i,t}\label{eq: gov_spending}\\
     &  T_{t} = \sum_{i \in N_{hh}}tax_{i, t}\label{eq: gov_taxes}\\
     &   B_{t} = B_{t-1} + G_{t} + UB_{t} + \bar{i}_{b}*B_{t-1} - T_{t} - Pcb_{t}\label{eq: gov_bills}\\
      &  D_{t} = \sum_{i \in N_{hh}}Dh_{i, t}\label{eq: bank_deposits}\\
    &  H_{t} = (\bar{\mu}_{cb} + \bar{v}) * D_{t}\label{eq: hpm}\\
    & Bb_{t} = \text{max}(D_{t} - L_{t} - H_{t}, 0)\label{eq: bank_bills}\\
    &  A_{t} = \left \{ \begin{array}{ll}
       H_{t} + L_{t} - D_{t}  &  \text{if}\; Bb_{t} = 0.0\;\\
       \bar{v} * D_{t}  & \text{otherwise}
    \end{array} \right . \label{eq: bank_advances}\\
    &  Pb_{t} = \bar{i}_{l}*\sum_{i \in N_{f}}L_{i, t-1} + \bar{i}_{b}*Bb_{t-1} + \bar{i}_{b}H_{t-1} - \bar{i}_{d}*D_{t-1}  - \bar{i}_{b}A_{t-1}\label{eq: bank_profits}\\
     & Bcb_{t} = B_{t} - Bb_{t}\label{eq: cb_bills}\\
    &  Pcb_{t} = \bar{i}_{b}Bcb_{t-1} - \bar{i}_{b}H_{t-1} + \bar{i}_{b}A_{t-1}\label{eq: cb_profits}\\
    &  c_{i, t} = \bar{\alpha}_{1}\frac{Yd_{i, t-1}}{p^{*}_{t}} + \bar{\alpha}_{2}\frac{Dh_{i,t-1}}{p^{*}_{t}}\label{eq: cons}\\
    &   C_{i,t} = c_{i,t} *p^{*}_{t} \label{eq: real_cons}\\
    &  tax_{i, t} = \bar{\tau}_{g} WB_{i,t-1}\label{eq: hh_taxes}\\
    &  Yd_{i, t} = WB_{i, t} + UB_{i, t} + \sum_{j \in N_{f}}Pf_{j,t} + Pb_{t} + \bar{i}_{d}*Dh_{i,t-1} - tax_{i,t}\label{eq: disp_income}\\
    & Dh_{i,t} = Dh_{i,t-1} + Yd_{i, t} - C_{i, t}\label{eq: hh_deposits}
\end{align}

\section{Additional tables}
\begin{table}[H]
    \centering
    \resizebox{\textwidth}{!}{\begin{tabular}{lccccccccc}
    \toprule
 &  & \multicolumn{2}{c}{Firms} & & \multicolumn{2}{c}{Bank} &  \multicolumn{2}{c}{Central Bank}&\\ \cmidrule{3-4} \cmidrule{6-7} \cmidrule{8-9}
& Government & Current & Capital & Households& Current & Capital & Current &Capital&$\Sigma$ \\ 
\toprule
    Consumption    & & $+C$ & & $-C$&  & & &&0 \\
    Government expenditures  & $-G$ & $+G$& & &  & & &&0  \\
    Change in inventories  & & $+\Delta INV$ &$-\Delta INV$ & &  & & &&0\\
    Wages  & & $-WB$& & $+WB$ &  & & &&0  \\ 
    Taxes  &$+T$ & & &$-T$ &  & & &&0  \\
    Unemployment benefits  &$-UB$ & & & $+UB$&  & &&& 0  \\ 
    \hline
    Interests on bills  & $-i_{b}B_{t-1}$& & &&$+i_{b}Bb_{t-1}$  & &$+i_{b}Bcb_{t-1}$&& 0  \\
  Interests on loans  & & $-i_{l}L_{t-1}$& & &  $+i_{l}L_{t-1}$& & &&0  \\
  Interests on reserves & & & &&$+i_{b}H_{t-1}$  & &$-i_{b}H_{t-1}$&& 0 \\
    Interests on advances & & & &&$-i_{b}A_{t-1}$  & &$+i_{b}A_{t-1}$&& 0 \\
    Interests on deposits  & & & &$+i_{d}Dh_{t-1}$ & $-i_{d}D_{t-1}$ & &&& 0  \\
Profits of firms  & &$-Pf$ & &$+Pf$ &  & & &&0  \\
Profits of banks  & & & & $+Pb$& $-Pb$ & & &&0  \\
Profits of cb  & $+Pcb$& & & &  && $-Pcb$& &0  \\

    \hline
$\Delta$bills  &$+\Delta B$ & & & && $- \Delta Bb$  && $-\Delta Bcb$& 0  \\
$\Delta$loans  & & & $+\Delta L$& && $-\Delta L$ & & &0  \\
$\Delta$deposits  & & & & $-\Delta Dh$&& $+\Delta D$  & && 0  \\
$\Delta$high powered money & & & & &&$-\Delta H$ && $+\Delta$H& 0 \\ 
$\Delta$advances & & & & &&$+\Delta A$ && $-\Delta$A& 0 \\ 
         \bottomrule
         $\Sigma$  & 0 & 0 & 0 & 0 & 0 & 0 &  0  \\
         \bottomrule
    \end{tabular}}
    \caption{Transaction flow matrix.}
    \label{tab: tfm}
\end{table}

\section{Initial values}
The SFC structure of the model allows the analytical computation of initial values after some assumptions are made on certain endogenous variables.
We start the simulations for each scenario assuming a full-employment condition, such that the total number of employees equals the total number of households $n = \bar{n}_{hh}$.
The wage amount $W$ has been assumed to be equal to the base wage $\omega_{b}$ (of FirmAsCAS model). After imposing an initial government debt to GDP ratio \citep{Lavoie.2019} and an initial stationary state, we could obtain the total amount of output $y$ produced by firms and all the other flows and stocks implied by the SFC structure. Also, we endogenously computed the initial value for $\alpha_{2}$ – households' propensity to consume out of deposits – such that the bills bought by the central bank equal the amount of reserves demanded and advances provided to banks.
These initial assumptions yielded a simplified system of equations from which the initial values of Table \ref{tab: initial_values} are derived.

\begin{table}[H]
    \centering
    \resizebox{!}{0.45\textwidth}{\begin{tabular}{llll}
    \toprule
       Symbol  & Remark & Description & Value \\
       \toprule
        $n_{hh}$ & free & Number of households& $500$ \\
        $n_{f}$ & free & Number of firms& $5$ \\
        $n_{b}$ & free & Number of banks& $1$ \\
        $n_{cb}$ & free & Number of central banks& $1$ \\
        $n_{g}$ & free & Number of governments & $1$ \\
        $n$ & pre-SS & Number of employed workers& $500$ \\
      $u$ & pre-SS & Number of unemployed households& $0$\\
      $y$ & pre-SS & Firms' output & $558.787517$\\
   
    $W$ & pre-SS & Wages & $8.0$\\
    $WB$ & SS-given & Wage bill & $4000.0$\\
    $UC$ & SS-given & Firms' unit costs& $7.158356$\\
   $s^{exp}$ & SS-given & Firms' expected sales& $558.787517$\\
    $inv$ & SS-given & Firms' real inventories& $558.787517$\\
    $INV$ & SS-given & Firms' inventories at current costs& $4000.0$\\
        $inv^{exp}$ & SS-given & Firms' expected real inventories& $558.787517$\\
    $inv^{T}$ & SS-given & Firms' inventories target& $558.787517$\\
    $p$ & SS-given & Firms' prices& $10.021698$\\
     $G$ & SS-given & Real government spending& $1002.16984$\\
     
       $c$ & SS-given & Households' consumption & $458.787517$\\
      $UB$ & SS-given & Unemployment benefits& $0.0$\\
      $T$ & SS-given & Taxes& $720.0$\\
      $C$ & SS-given & Households' real consumption & $ 4597.830153$\\
      $Yd$ & SS-given & Households' disposable income & $4597.830153$\\
      $D_{h}$ & SS-given & Households' deposits & $4502.908765$\\
      $D$ & SS-given & Bank's deposits & $4502.908765$\\
      $L$ & SS-given & Firms' loans & $4000.0$\\
    $P_{f}$ & SS-given & Firms' profits & $1580.0$\\
    $H$ & SS-given & Bank's reserves & $1576.018067$\\
    $B_{b}$ & SS-given & Bank's bills & $0.0$\\
    $A$ & SS-given & Bank's advances & $1073.109302$\\
     $P_{b}$ & SS-given & Bank's profits & $8.5029087$\\
     $B_{cb}$ & SS-given & Central bank's bills & $502.9087653$\\
     $P_{cb}$ & SS-given & Central bank's profits & $0.0$\\
  $B$ & SS-given & Government's bills & $502.9087653$\\

        \bottomrule
    \end{tabular}}
    \caption{Initial values of the aggregate sectors.}
    \label{tab: initial_values}
\end{table}
\subsubsection{Derivation of output}
By assuming that at the stationary state $\frac{\Delta B}{y_{t}} = 0$ and that $B_{t-1} = B_{t}$, we have the following expressions.
\begin{align*}
    &\frac{G_{t} + UB_{t} - T_{t} - Pcb_{t} + \bar{i}_bB_{t-1}}{y_{t}} = 0 \tag{a}\\
    &\frac{\bar{g}*\sum_{i\in N_{f}}p_{i,t} + UB_{t} - T_{t} - Pcb_{t}}{y_{t}} + \bar{i}_{b}\frac{B_{t}}{y_{t}} = 0 \tag{b}
\end{align*}
By imposing that $\frac{B_{t}}{y_{t}} = \bar{b}_{y}$, we thus have:
\begin{align*}
      y_{t}^{*} = \frac{(\sfrac{\bar{g}  * W)}{} {\bar{pr} - Pcb_{t}}}{ \sfrac{(\bar{\tau}_{g} * W)}{\bar{pr}} - \bar{i}_b * \bar{r}} \tag{c}
\end{align*}

\subsection{Additional figures}
\begin{figure}[H]
\centering
    \includegraphics[scale = 0.53]{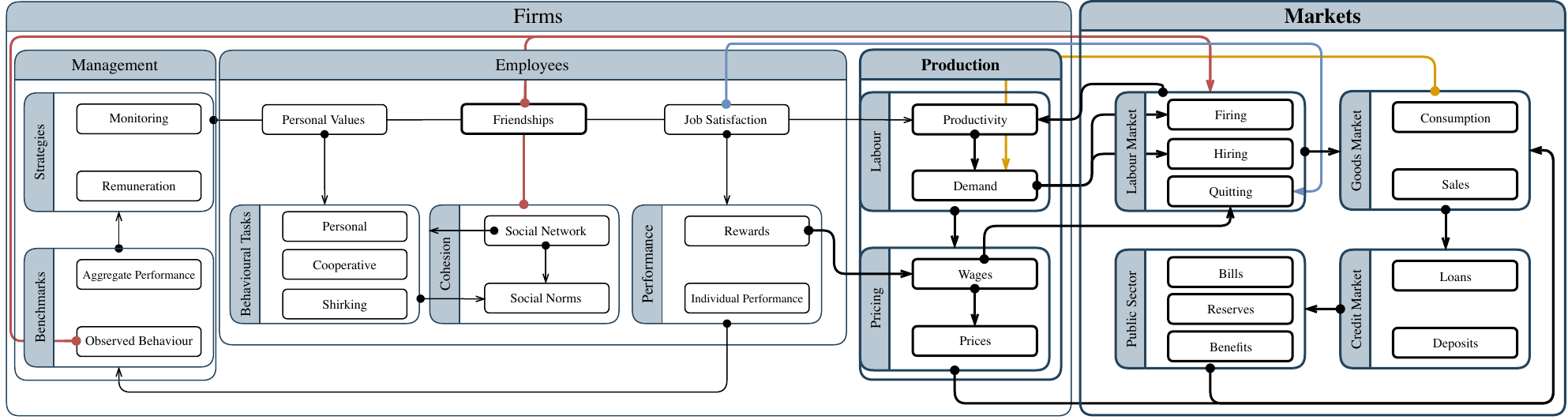}
    \caption{Detailed model overview, extension of \cite{Banning.2023}. Authors' own elaboration. Elements in bold indicate the newly added features.}
    \label{fig:model-overview}
\end{figure}

\end{document}